\documentclass[aps,twocolumn,amsmath,amssymb,amsfonts,floatfix]{revtex4-1}
\usepackage{epsfig}
\usepackage{epstopdf}
\usepackage{ifpdf} 
\usepackage{graphicx}
\usepackage{color}
\usepackage[usenames,dvipsnames]{xcolor}
\usepackage{bm} 
\usepackage{hyperref}
\usepackage[normalem]{ulem}
\usepackage[english]{babel}
\usepackage{natbib}
\usepackage{bbold}
\usepackage{verbatim}
\usepackage{floatrow}
\usepackage[caption=false]{subfig}
\begin{document}
\title{Effective interactions mediated between two permeable disks in an active fluid}

\author{Mahmoud Sebtosheikh}
\thanks{mahmoud-sebtosheikh@ipm.ir (corresponding author)}
\affiliation{School of Physics, Institute for Research in Fundamental Sciences (IPM), P.O. Box 19395-5531, Tehran, Iran}
\author{Ali Naji}
\thanks{a.naji@ipm.ir}
\affiliation{School of Physics, Institute for Research in Fundamental Sciences (IPM), P.O. Box 19395-5531, Tehran, Iran}

\begin{abstract}
We study steady-state properties of a bath of active Brownian particles (ABPs) in two dimensions in the presence of two fixed, permeable (hollow) disklike inclusions, whose interior and exterior regions can exhibit   mismatching motility (self-propulsion) strengths  for the ABPs. We show that such a discontinuous  motility field strongly affects spatial distribution of ABPs and thus also the effective interaction mediated between the inclusions through the active bath. Such net interactions arise from soft interfacial repulsions between ABPs that sterically interact with and/or pass through permeable membranes assumed to enclose the inclusions. Both regimes of repulsion and attractive (albeit with different mechanisms) are reported and summarized in overall phase diagrams. 
\end{abstract}

\maketitle

\section{Introduction}
\label{sec:intro}

Effective interactions between colloidal inclusions in fluid media  is of crucial importance in  predicting phase behaviors of colloidal suspensions and manipulating them for technological applications. Depletion forces are one of the primary types of effective interactions that emerge in size-asymmetric suspensions of (large) inclusions and (small) depletant particles \cite{Lekkerkerker2011,Likos2001}. They arise due to steric exclusion of depletants from the proximity of inclusions, are typically short-ranged and attractive. 

Even though depletion forces are thoroughly investigated in nonactive systems \cite{Lekkerkerker2011}, their nonequilibrium counterparts, especially in the context of active Brownian particles (ABPs) \cite{revLowen2016,revGompper2015,revStark2016, revMarchetti2013}, have been studied only in the recent past  \cite{revSpeck2020,Harder2014,Naji2017,Lowen2015,Cacciuto2014,Bolhuis2015,Leonardo2011,Reichhardt2014,Ferreira2016,Narayanan2018,Zhang2018,Zhang2020, Pagonabarraga2019,Kafri2018, Selinger2018, Garcia2015, Narayanan2018, Naji2019, Yang2020, Marconi2018, Mishra2018}. Unlike nonactive particles, ABPs strongly accumulate near rigid boundaries due to their persistent  motion  (see, e.g., Refs. \cite{Lowen2008,Gompper2013,Mallory2014}). This leads to near-wall layering of ABPs  \cite{Harder2014,Naji2017,Bolhuis2015,Zhang2018,Zhang2020, Pagonabarraga2019,Naji2019} due to the counterbalancing steric repulsions between them in the highly populated near-surface regions, which in turn causes qualitative changes in the effective interactions mediated between inclusions in the active bath. In the particular example of hard disklike  inclusions that will be of interest here, the effective interaction between two such inclusions fixed within a two-dimensional  active bath was shown to be predominantly {\em repulsive}  \cite{Harder2014,Bolhuis2015,Ferreira2016,Yang2020}, featuring nonmonotonic distance-dependent behaviors due to the mentioned ABP layering, or ring formation around the disks \cite{Naji2017,Pagonabarraga2019,Yang2020,Naji2019}. Such active bath-mediated interactions have been investigated in different cases, elucidating their dependence on the geometric shape and relative size of the inclusions \cite{Harder2014,Ferreira2016,Zhang2018,Zhang2020,Reichhardt2014,Bolhuis2015} as well as motility strength, concentration  \cite{Bolhuis2015,Ferreira2016,Zhang2018,Zhang2020,Yang2020,Narayanan2018} and chirality of ABPs \cite{Naji2017}. In certain cases, attractive interactions \cite{Zhang2018,Zhang2020,Pagonabarraga2019,Naji2017,Naji2019,Reichhardt2014,Bolhuis2015} and noncentral  forces \cite{Pagonabarraga2019} have been reported. While the focus has mainly been on two-body interactions between fixed inclusions, other cases with mobile inclusions \cite{Yang2020, Mishra2018,Garcia2015,Harder2014,Leonardo2011,Kafri2018,Narayanan2018} and many-body mixtures  \cite{Mishra2018}  have also been studied, and importance of inclusion mobility \cite{Yang2020} and many-body interactions \cite{Naji2017} have been noted. 

While previous studies have explored various aspects of bath-mediated interactions in the case of hard (impenetrable) inclusions, possible implications of active-particle penetration into soft or hollow inclusions have not been considered in the said context. Permeable inclusions may be realized by soft micro-compartments and vesicles such as  liposomes and polymersomes \cite{Larsson2006,Kamat2011,Rideau2018}, immiscible/stabilized droplets in emulsions as well as active droplets \cite{Cates2017,Tjhung2012,Hyman2014,Julicher2017,Naji2018}. These structures can be utilized in a wide range of applications such as controlled internalization of polystyrene microparticles and crystallites in fibroblast cells \cite{Kodali2007}, selective entrapment of catalytically active nanoparticles within polymer stomatocytes \cite{Wilson2012}, targeted cargo/drug delivery to  soft tissues (such as tumors) by means of nonactive and active particles \cite{Peng2018,Joseph2017,Schmidt2018,Sitti2019,Magdanz2019,Ghosh2020} as well as biological microswimmers  \cite{Felfoul2016,Ghosh2020}. Despite such emerging applications, detailed modeling of active-particle penetration through flexible membranes have been considered only recently \cite{Daddi2019a,Daddi2019b}, with a growing number of studies having been focused on structural and dynamical ramifications of entrapping active particles within soft semipermeable vesicles, whose deformable  enclosing membranes are permeable only to the background fluid and not the active particles \cite{Tian2017, Chen2017, Shan2019, Wang2019, Paoluzzi2016, Spellings2015} (see also Ref. \cite{Vladescu2014} for an experimental study of entrapping motile bacteria within an emulsions drop). 

Here, we aim at investigating  spatial distribution of model ABPs and effective interactions mediated by them between two {\em permeable} dislike inclusions placed at fixed positions in a  two-dimensional active bath. The inclusions represent hollow fluid enclosures with permeating membranes defined via soft, repulsive interfacial potentials set in ring-shaped regions around them. This enables us to address the role of an inhomogeneous motility field (see, e.g.,  Refs. \cite{Schnitzer1990,Schnitzer1993,Brendel2015, Lowen2016, Lowen2018, Cate2016, Brader2017,Lowen2019,Merlitz2018}) by taking different self-propulsion strengths for ABPs inside and outside the inclusions. In general, ABPs display larger concentrations in regions of lower motility due to their locally  increased persistence times. Ordered states such as crystal and liquid crystalline  phases of ABPs have thus been found in low-motility media \cite{Brendel2015,Lowen2018}, and self-assembled structures such as membranes of  self-propelled rods have been reported to form spontaneously  at the interface of two media, exhibiting mismatching motility strengths \cite{Lowen2018}. We analyze various implications of the assumed inclusion permeability   and the motility-field discontinuity, and  determine the regimes of repulsion and attraction between the inclusions that emerge over the parameter space spanned by the interior/exterior P\'eclet numbers. 

\section{Model and Methods}
\label{sec:model}

\begin{figure}[t!]
\begin{center}
\includegraphics[width=0.85\textwidth]{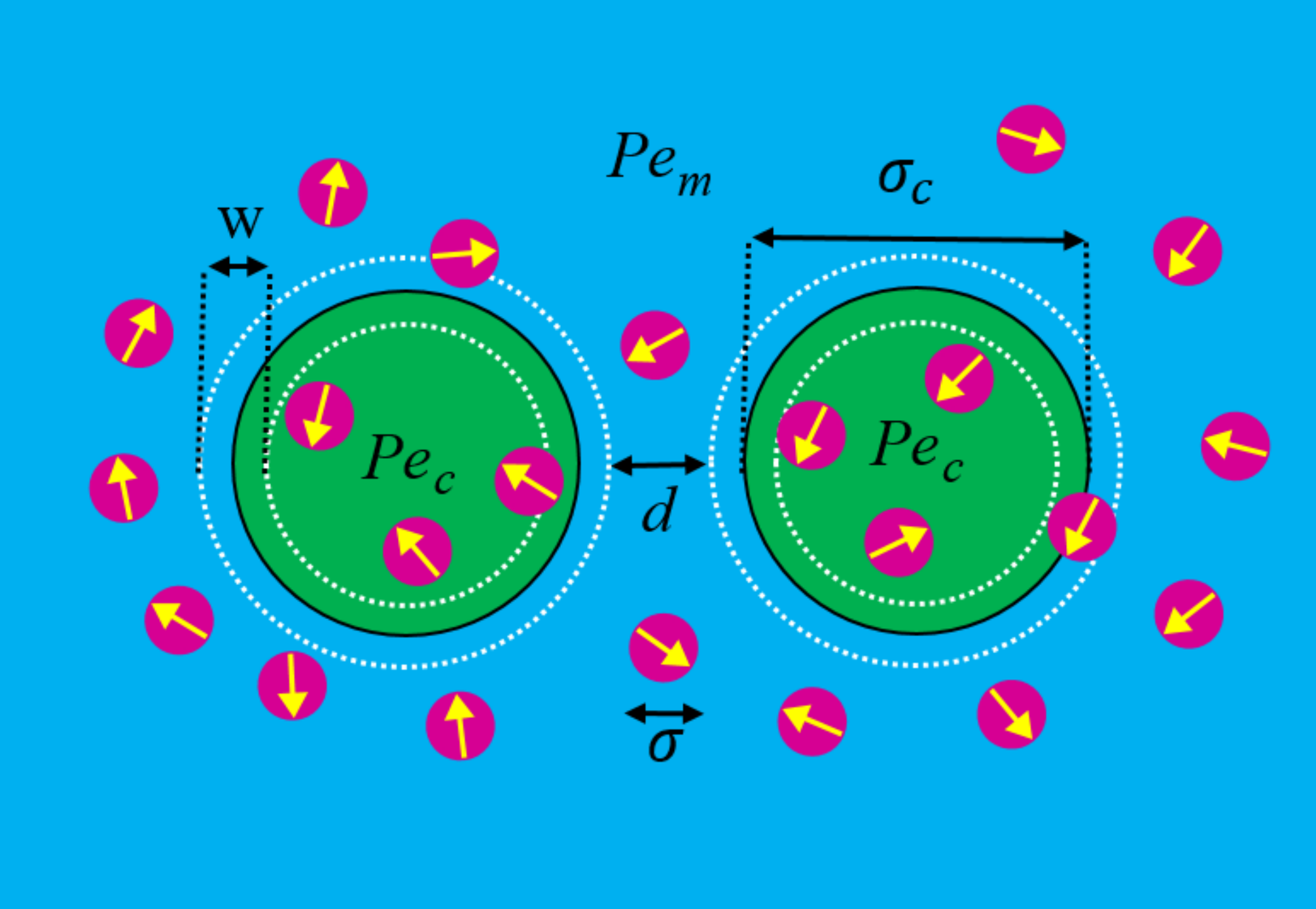}
\caption{Schematic view of two fixed permeable inclusions in a bath of active Brownian particles (of diameter $\sigma$) with mismatching P\'eclet numbers  inside ($Pe_c$) and outside ($Pe_m$) the inclusions. White dashed  circles show the inner/outer borders (of diameters $\sigma_c\mp w$) and the midlines of permeable  membranes (of thickness $w$) that enclose each inclusion. }
\label{fig:schematic}
\end{center}
\end{figure}

We consider a two-dimensional minimal model of disklike active Brownian particles (ABPs)   \cite{revLowen2016} with constant self-propulsion speeds and  diameter $\sigma$ in a base fluid, involving two fixed fluid enclosures, or hollow disklike inclusions of effective diameter $\sigma_c$; see Fig. \ref{fig:schematic}. Each inclusion  has a permeable  enclosing membrane of thickness $w=\sigma$, modeled by a soft steric potential (see below) that acts within the annulus of inner/outer radii $(\sigma_c\mp w)/2$. The  inclusion centers are at $\mathbf{R}_{1}$ and $\mathbf{R}_{2}$ along the $x$-axis at outer surface-to-surface distance $d$ (we label  the inclusion on the left/right by $k=1,2$, respectively). 

The ABPs exhibit different self-propulsion speeds, $v_c$ and $v_m$, inside and outside the inclusions, respectively, expressed through the discontinuous motility field \cite{Brendel2015,Lowen2018}
\begin{equation}
v({\mathbf r})=v_c+(v_m-v_c)\!\!\sum_{k=1,2}\!\!\Theta\!\left(|{\mathbf r}-\mathbf{R}_k|-\sigma_{c}/2\right),
\end{equation} 
with ${\mathbf r}=(x,y)$ being the spatial coordinates and $\Theta(\cdot)$  the Heaviside step function. The ABPs (labeled by $i=1\ldots,N$) are characterized by their position and self-propulsion orientation vectors   $\{{\mathbf r}_i(t)\}=\{(x_i(t), y_i(t))\}$  and $\{{\mathbf{n}_i (t)\}=\{(\cos\theta_i(t),\sin\theta_i(t))}\}$, respectively. Their stochastic  dynamics are given by the Langevin equations 
\begin{equation}
\label{Alang1}
\dot{\mathbf r}_i(t)=v\left({\mathbf r}_i(t)\right)\mathbf{n}_i(t)-\mu_{T}\frac{\partial U}{\partial{\mathbf r}_i}+\mathbf{\eta}_i(t);\,\,\,\, \dot{\theta_i}(t)=\zeta_i(t), 
\end{equation}
with angular coordinates $\{\theta_i\}$ measured from the $x$-axis, $\mu_T$ being the translational mobility, $U=U(\{{\mathbf r}_j\},\{\mathbf{R}_k\})$ the sum of interaction potentials between the constituent particles (see below),  and $\mathbf{\eta}_i(t)$ and $\mathbf{\zeta}_i(t)$ being translational and rotational white noises, respectively. The noise distributions are assumed to be Gaussian with zero mean, $\langle \mathbf{\eta}_i(t) \rangle=\langle \mathbf{\zeta}_i(t) \rangle=0$, and correlators $\langle \mathbf{\eta}_i(t) \mathbf{\eta}_j(t')\rangle=2D_{T}\delta_{ij}\delta(t-t')$ and $\langle\mathbf{\zeta}_i(t)\mathbf{\zeta}_j(t')\rangle=2D_{R}\delta_{ij}\delta(t-t')$, where $D_T$ and $D_R$ are the translational and rotational diffusivities, respectively. We adopt the Einstein-Smoluchowski-Sutherland relation, $D_{T}=\mu_{T}k_{\mathrm{B}}T$, and Stokes diffusivities for the no-slip spheres, yielding  $D_{R}={3D_{T}}/{\sigma^{2}}$ \cite{Brenner1983}. 

The ABPs are assumed to interact via a Weeks-Chandler-Andersen (WCA) pair potential, giving the  energy of interaction between the $i$th and the $j$th ABP as
\begin{eqnarray}
\!\!\!U_{\textrm{WCA}}^{(ij)}\!=\!\left\lbrace 
\begin{array}{ll}
\!4\epsilon\bigg[\!\!\left(\frac{\sigma}{r_{ij}}\right)^{12}\!\!-\left(\frac{\sigma}{r_{ij}}\right)^{6}\!\!+\!\frac{1}{4}\bigg]
\,\,&: \,\,r_{ij}\leq 2^{1/6}\sigma,\\ 
\!0 \,\,&: \,\, r_{ij}> 2^{1/6}\sigma,
\end{array}
\right.
\label{WCA}
\end{eqnarray}
where $r_{ij}=|{\mathbf r}_i-{\mathbf r}_j|$. As they go through the interfacial regions (membranes) of the inclusions, the ABPs  experience a soft, repulsive, WCA-type potential,  which we adopt from Ref.  \cite{Abkenar2013}, giving the energy of  interaction between the $i$th ABP and the $k$th inclusion as
\begin{equation}
U_{\textrm{sWCA}}^{(ik)}\!=\!\left\lbrace
\begin{array}{ll}
\!\!4\epsilon'\bigg[\!\!\left(\!\frac{\sigma'}{\sqrt{{r'}_{ik}^2+\alpha^{2}}}\!\right)^{12}\!\!\!\!\!-\!\left(\!\!\frac{\sigma'}{\sqrt{{r'}_{ik}^2+\alpha^{2}}}\!\right)^{6}\bigg]\!\!+\!U_{0} &:r'_{ik}\leq\sigma'\!, \\
\!0 &:r'_{ik}> \sigma'\!, 
\end{array}
\right.
\label{SWCA}
\end{equation}
with $r'_{ik}\!=\!\big||{\mathbf r}_i-\mathbf{R}_k|-\sigma_c/2\big|$,    $\alpha^2\!=\!(2^{1/3}-1)\sigma'^2$, $2\sigma'\!=\!\sigma+w$, $U_0=-4\epsilon'\lambda^3(\lambda^3-1)$, and $\lambda^{-1}=1+(\alpha/\sigma')^{2}$. 
We use representative values of $\epsilon=10k_{\mathrm{B}}T$ and $\epsilon'=0.0127k_{\mathrm{B}}T$. 

\begin{figure*}
\begin{center}
\includegraphics[width=\textwidth]{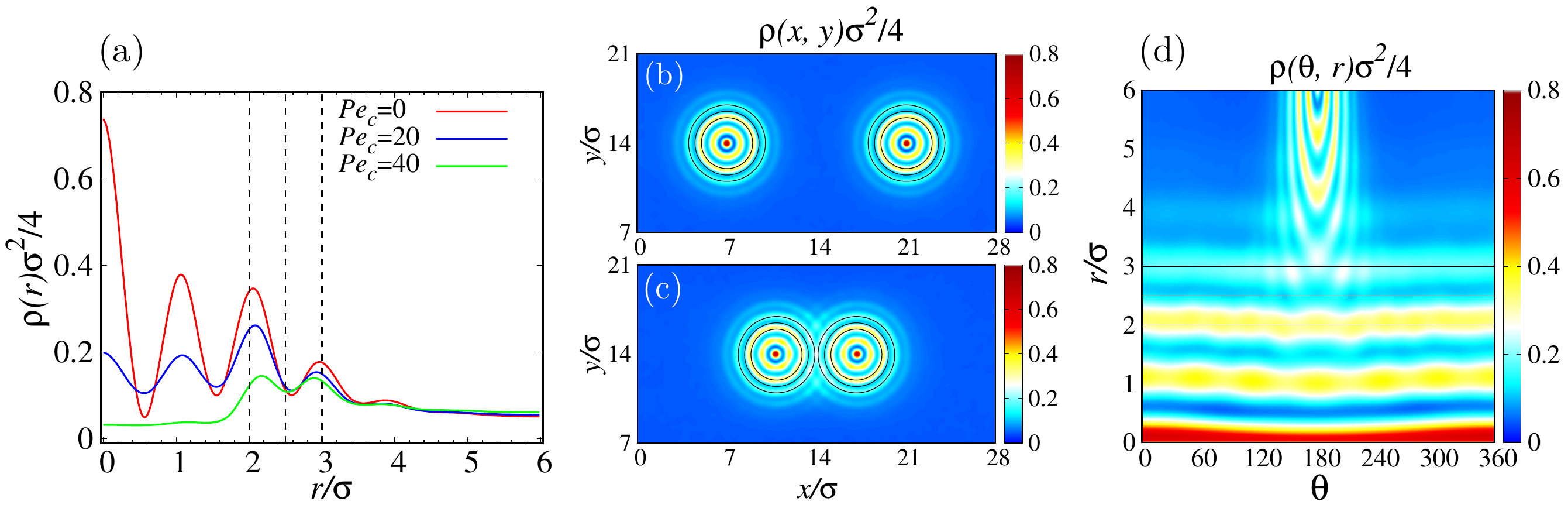}
\caption{(a) Rescaled radial density profiles of ABPs as functions of the rescaled radial distance from the center of an {\em isolated} inclusion for  $Pe_{c}=0,20,40$ and fixed $Pe_{m}=40$. (b) and (c) show color-coded density maps of ABPs in and around the inclusions   at surface-to-surface distances $d/\sigma=8$ and 0.125, respectively, and fixed $Pe_c=0$, $Pe_m=40$, and $\phi=0.2$. (d) Same as (c) but in polar coordinates with the origin being taken at  the center of the right inclusion. In (b)-(d) red/yellow colors indicate high/intermediate particle densities, while cyan/blue indicate reduced/nearly vanishing particle densities. The interfacial region around the inclusions is indicated  by three black solid circles (b, c) or horizontal lines (d); see also Fig. \ref{fig:schematic}. 
} 
\label{fig:fig2}
\end{center}
\end{figure*}

The Langevin equations \eqref{Alang1} are numerically solved using standard Brownian Dynamics simulations, and the behavior of the system will be characterized in terms of the key dimensionless parameters given by the  interior ($c$) and exterior ($m$) P\'eclet numbers,
\begin{eqnarray}
\label{Pe}
Pe_{c,m}=\frac{\sigma\,v_{c,m}}{2D_{T}}, 
\end{eqnarray}
the rescaled center-to-center distance of the inclusions, $d/\sigma$, the rescaled inclusion size, $\sigma_c/\sigma$, and the area fraction of ABPs, $\phi$. In the simulations, we use $N=200-800$ ABPs at fixed  area fractions $\phi=\{0.2, 0.4\}$, with $\phi= N\pi\sigma^2/(4L_xL_y)$ defined in the absence of the inclusions. $L_x$ and $L_y$ are lateral dimensions of the simulation box with periodic boundary conditions. Strong correlations of ABPs with inclusion boundaries suppress motility-induced clustering/phase separation of ABPs \cite{Baskaran2013} that can  otherwise arise in the bulk.  We fix the inclusion size as $\sigma_c=\{5\sigma, 10\sigma\}$ with $L_x=L_y=\{28\sigma, 56\sigma\}$, respectively (we use $L_x=2L_y=56\sigma$ in Fig. \ref{fig:fig8}(b) to capture the full range of the interaction force without undesired boundary effects). The area fraction of the inclusions $\phi_c= \pi\sigma_{c}^2/(2L_xL_y)=0.05$ ($0.025$ in Fig. \ref{fig:fig8}(b)). The simulations are performed for $(2-6)\times 10^{7}$ timesteps of rescaled size $D_T\delta t/\sigma^2=1.33\times 10^{-5}$. The first $10^{7}$ steps are used for relaxation purposes, and the steady-state averages are calculated over 3-20 independent simulations.

\section{Results}

\subsection{Spatial distribution of ABPs}
\label{sec:spatial}

When the inclusions are placed at relatively large surface separations, the spatial distribution of ABPs around each   of them exhibits radial symmetry, with a typical  radial number density profile, $\rho (r)$, from the given inclusion center as shown in Fig. \ref{fig:fig2}(a) for fixed $d/\sigma=8$, $Pe_m=40$ and different $Pe_c$. As seen, for $Pe_c=0$ (nonactive interior; red solid curve), there is a larger concentration of ABPs inside the inclusions relative to the cases with $Pe_c>0$ (blue and green solid curves), in accord with previous findings in inhomogeneous systems \cite{Schnitzer1990,Schnitzer1993,Brendel2015}. The oscillatory behavior of $\rho (r)$  reflects ringlike ABP layers, forming due to steric repulsions between the ABPs inside the inclusions, as depicted also in  Fig. \ref{fig:fig2}(b). As $Pe_c$ is increased, ABPs at the central regions inside the inclusions are more strongly depleted than those near the interfacial regions ($2\leq r/\sigma\leq 3$), as ABPs are slowed down by the enclosing membrane potential and, thus, create larger steady-state densities near the interfacial regions. 

The radial symmetry of the ABP distribution is broken when the two inclusions are placed nearby. This is shown in Fig. \ref{fig:fig2}(c), where multiple intersections between individual rings formed around each inclusion are  illustrated; they  are discerned more clearly in the closeup view in Fig. \ref{fig:fig2}(d). It is this asymmetric distribution of ABPs in and around the inclusions that causes an effective interaction force between them, which we shall explore in the following section.

\begin{figure}[t!]
\begin{center}
\includegraphics[height=5.5cm]{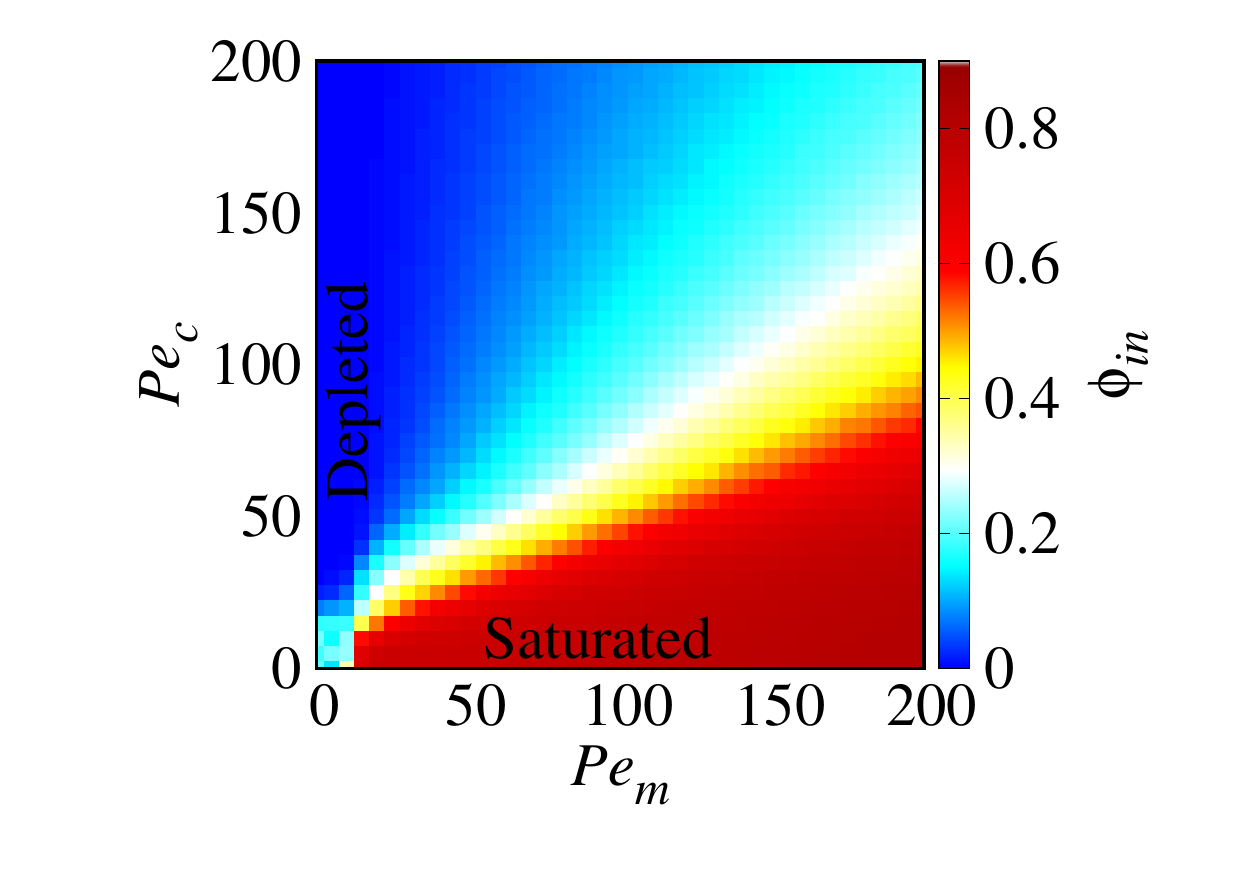}
\caption{Internal area fraction of ABPs within the permeable inclusions as a function of $Pe_c$ and $Pe_m$ for $\phi=0.2$ and $\phi_c=0.05$, when the inclusions are at contact, $d=0$.}
\label{fig:fig3}
\end{center}
\end{figure}

Before proceeding further, we briefly examine the {\em internal} area fraction of ABPs within the inclusions as a function of  $Pe_c$ and $Pe_m$; see Fig. \ref{fig:fig3}. Being defined as $\phi_{in}={N_{in}\sigma^{2}}/{(2\sigma_c^2)}$, where $N_{in}$ is the number of particles trapped inside both inclusions, $\phi_{in}$ increases by increasing $Pe_m$ or decreasing $Pe_c$, which, as noted before, is due to the relatively longer times ABPs spend inside the inclusions. However, $\phi_{in}$ cannot exceed a certain value due to steric repulsions between ABPs \cite{Brendel2015,Cate2016}, leading to a finite saturation level (dark red colors in the figure). 

\subsection{Effective interactions and force-distance profiles}
\label{subsec:fprofiles}

Due to the up-down symmetry of the system, only the $x$-component (along the center-to-center line) of the effective force acting on the inclusion is nonzero. It can be obtained, upon appropriate steady-state averaging, from the simulations using 
\begin{equation}
\label{F}
F(d)=\sum_{i=1}^N\bigg\langle\frac{\partial}{\partial {x_i}}U_{\textrm{sWCA}}^{(i2)}\bigg\rangle, 
\end{equation} 
where the expression inside the brackets is the $x$-component of the instantaneous force imparted by the $i$th ABP on the   $k=2$ inclusion  (on the right in Fig. \ref{fig:schematic}) and $U_{\textrm{sWCA}}^{(ik)}$ is defined by  Eq. \eqref{SWCA}. Evidently, $F(d)>0$ ($F(d)<0$) represents a repulsive (attractive) interaction force. We find identical results for the averaged force on the left inclusion within the simulation error bars. 

Figure \ref{fig:fig4}(a) shows the effective force between the permeable inclusions as a function of their surface-to-surface distance, $d$, at fixed exterior P\'eclet number, $Pe_m=20$, and for the interior P\'eclet number increased from $Pe_c=0$ up to $Pe_c=160$. As a general trend, the force profiles indicate repulsive interactions with a characteristic oscillatory behavior, exhibiting  a number of successive local maxima and minima,  as the overall magnitude of the force decreases with $d$. These local minima and maxima can be understood directly based on the overlaps between the rings of ABPs that form around each of the inclusions (and also the intersections of the rings associated with one inclusion with the bounding surface of the other inclusion) as $d$ is varied. These mechanisms have been elucidated in detail in Ref. \cite{Naji2017}. As seen in Figure \ref{fig:fig4}(b), a similar behavior is found for the effective force, when the interior P\'eclet number, $Pe_c$, is kept fixed (here, $Pe_c=0$) and $Pe_m$ is varied. The dependence of the force profiles in the two above-mentioned cases, however, turns out to be distinctly different, as we shall explore next. 

\subsection{Role of interior/exterior motility strengths}
\label{subsec:inout_peclet}

\begin{figure}[t!]
\begin{center}
\includegraphics[width=0.725\textwidth]{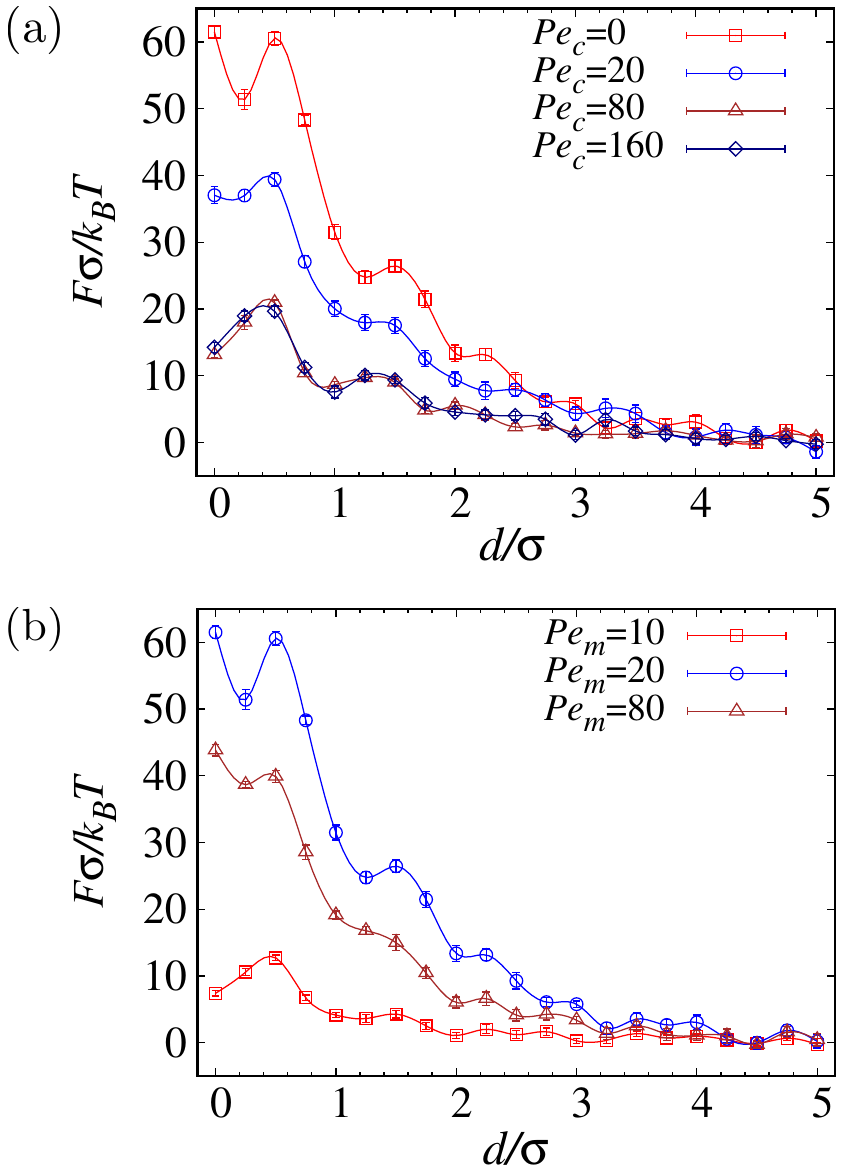}
\caption{Effective force acting on the inclusions as a function of their rescaled surface-to-surface distance, $d/\sigma$, for (a)  fixed $Pe_m=20$ and different values of $Pe_c$, and (b) fixed $Pe_c=0$ and different values of $Pe_m$, as indicated on the graph. In both panels, $\phi=0.2$ and $\phi_c=0.05$.}
\label{fig:fig4} 
\end{center}
\end{figure}

\begin{figure}[t!]
\begin{center}
\includegraphics[width=0.75\textwidth]{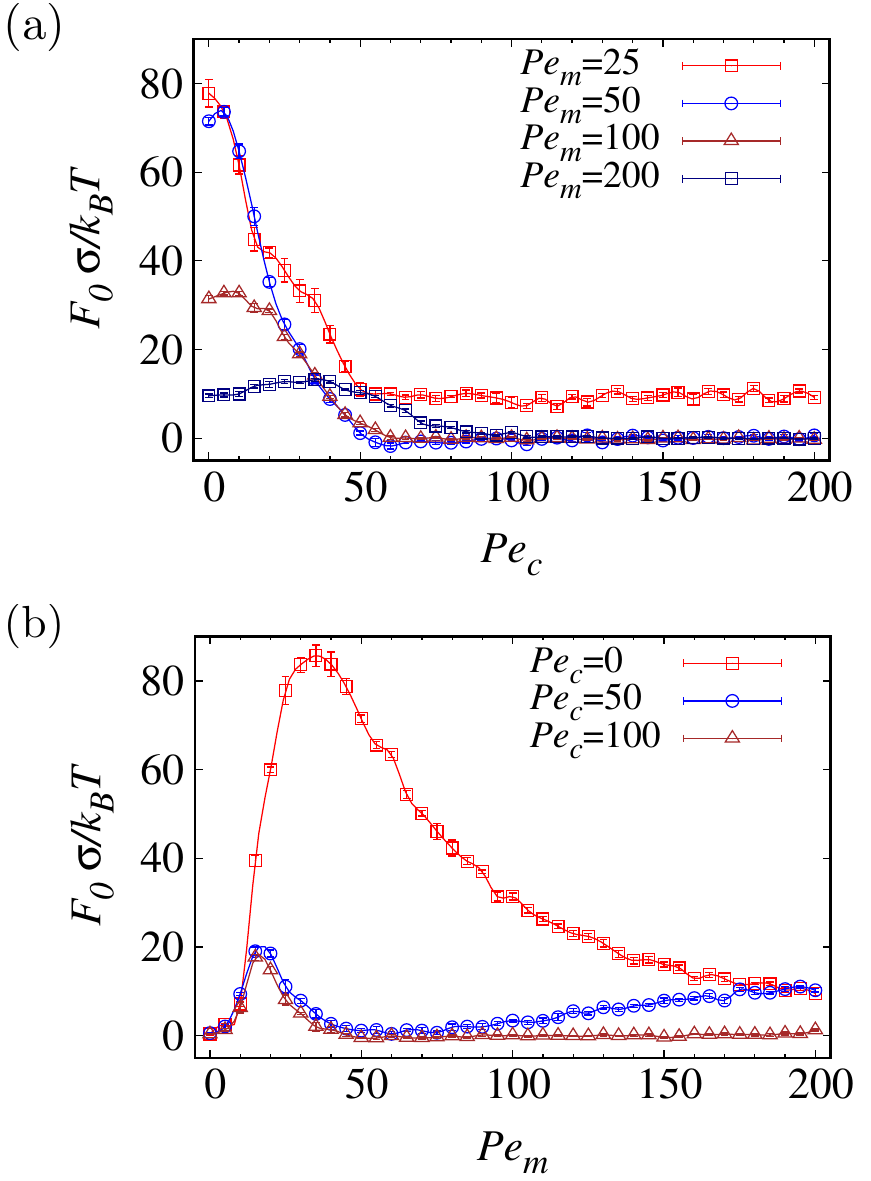}
\caption{(a) Contact force, $F_0=F(d=0)$, on the inclusions as a function of $Pe_c$ for different fixed values of $Pe_m$,  and (b) as a function of $Pe_m$ for different fixed values of $Pe_c$, as shown on the graphs. In both panels, $\phi=0.2$ and $\phi_c=0.05$.}
\label{fig:fig5}
\end{center}
\end{figure}

Our data in Fig. \ref{fig:fig4} show that, as the interior P\'eclet number, $Pe_c$, is increased at fixed exterior P\'eclet number, $Pe_m$, the magnitude of the force acting on the inclusions  decreases, converging to a limiting curve already for $Pe_c=80$. This behavior is illustrated in Fig. \ref{fig:fig5} (top), where we concentrate on the contact force $F_0=F(d=0)$ as a function of $Pe_c$ at different fixed values of $Pe_m$. The interaction force on the inclusions drops smoothly as $Pe_c$ increases. When $Pe_m$ is small  (see, e.g., the data with  $Pe_m=25$ in Fig. \ref{fig:fig5}, or $Pe_m=20$ in Fig. \ref{fig:fig4}(a)), the force decreases monotonically and tends to a nonvanishing constant as $Pe_c$ is increased. When $Pe_m$ is sufficiently large (e.g., $Pe_m=50$), the force behavior with $Pe_c$ exhibits a weak and broad maximum before it drops to zero. This behavior can be understood as follows.

The effective force acting on the inclusions can be viewed as a resultant effect of two separate force components  imparted by interior and exterior ABPs on the enclosing membrane of the inclusions, to be referred to as internal and external forces, respectively. To  understand the dependence of the effective force on particle motility, one needs to understand how these force components vary with the P\'eclet numbers. The force components are directly influenced by the concentration of ABPs accumulated near the boundary regions of the inclusions and, specifically, also within their interior regions. 

In the cases where  $Pe_m$ is fixed  (Fig. \ref{fig:fig5}(a)), the external force depends only on the effective hardness of the inclusions  and is  directly related to the internal ABP concentration. It is nevertheless important to note that in addition to the hardness produced by the enclosing  membrane, external ABPs also encounter resistance from internal ABPs accumulated within the inclusions. At different values of fixed $Pe_m$, by increasing $Pe_c$, external force monotonically decreases down to a constant value. This is because, when the interior regions are nearly fully depleted from ABPs (see Fig. \ref{fig:fig3}), effective hardness of the inclusions reduces to that of the enclosing membranes. By increasing $Pe_c$, the internal force does not remain constant. The internal force influenced by ABPs distribution near the enclosing membrane inside the inclusions which depends on the velocity and the concentration of internal ABP and, as such, internal force varies proportionally with these quantities. At low $Pe_m$ such as $Pe_m=25$, the internal ABP concentration rapidly decreases and, hence, the internal force monotonically decreases. At high $Pe_m$ such as $Pe_m=200$, however, as $Pe_c$ is initially increased, the internal ABP concentration remains constant then decreases slowly; hence, the internal force increases up to a maximum, because of the constant concentration and the increase in typical velocities of the ABPs that enables them to penetrate more strongly within the membranes enclosing the inclusions. The internal force decreases beyond the maximum because of a decreased ABP concentration.

\begin{figure*}[t!]
\begin{center}
\includegraphics[width=\textwidth]{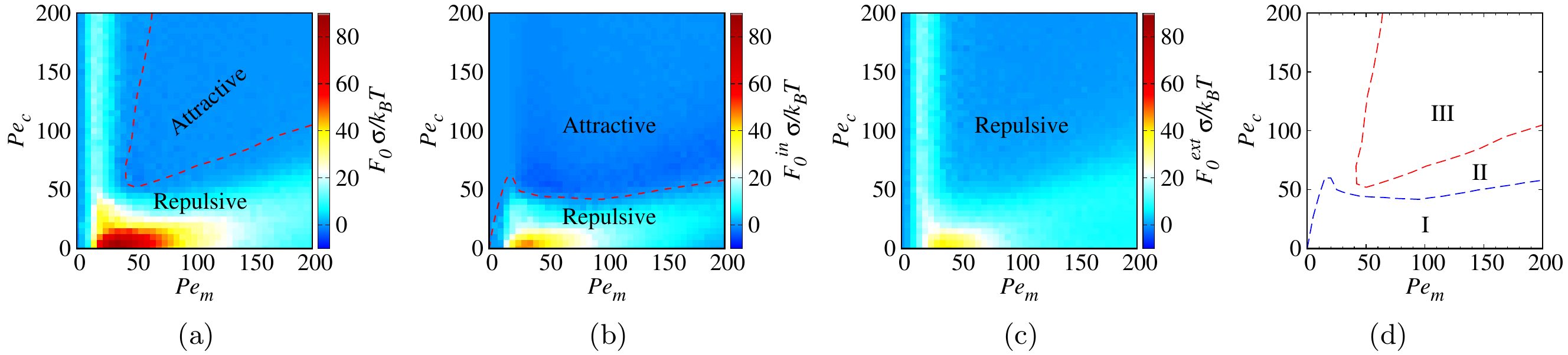}	
\caption{(a) The net effective force acting on the inclusions at contact plotted along with (b) the internal force and (c) the external force components as functions of P\'eclet numbers $Pe_c$ and $Pe_m$ for $\sigma_c=5$, $\phi=0.2$ and $\phi_c=0.05$. The dashed curves in (a) and (b)  show the borders between repulsive and attractive force regimes. (d) Borders of different parametric regions I-III   determined based on the ABP distribution and the force components as discussed in the text.}
\label{fig:fig6}
\end{center}
\end{figure*}

\begin{figure*}[t!]
\begin{center}
\includegraphics[width=0.75\textwidth]{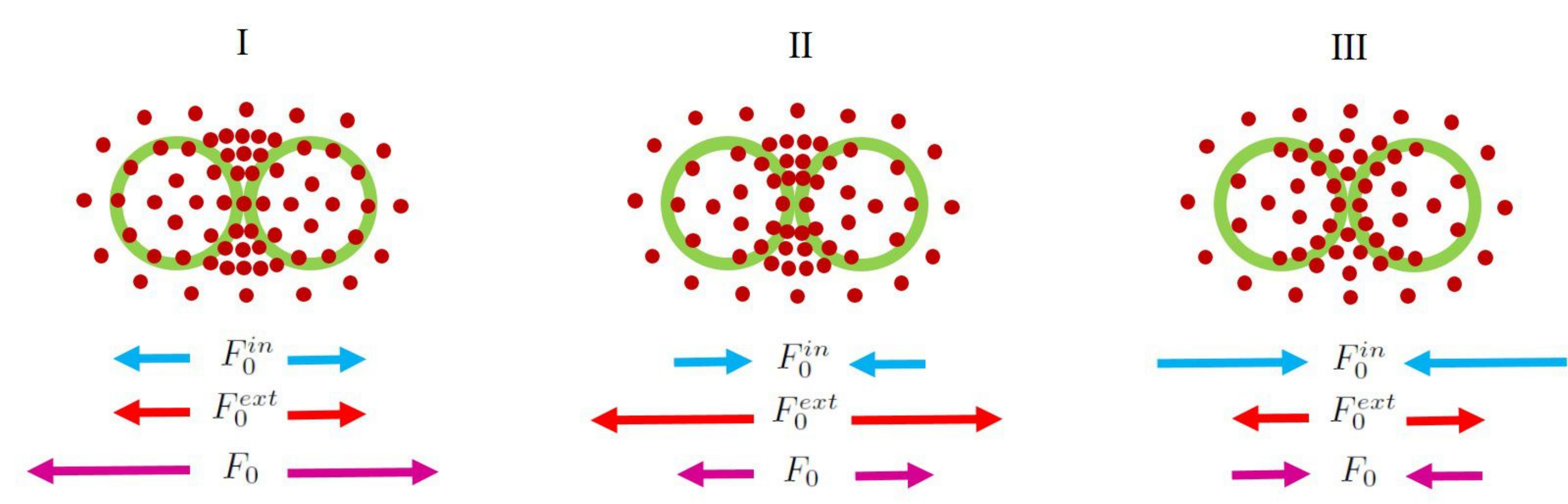}
\caption{Mechanisms of force generation in regions I-III (see Fig. \ref{fig:fig6} and the text). Because of the preferred spatial distribution of ABPs, the external component of the force  on the inclusions remains always repulsive, while the internal component can become repulsive (I) or attractive (II, III), giving rise to overall repulsion (I, II) or attraction (III) between the inclusions.}
\label{fig:fig7} 
\end{center}
\end{figure*}

We now turn to the case, where  the exterior P\'eclet number, $Pe_m$, is varied at fixed interior P\'eclet number, $Pe_c$. Figure \ref{fig:fig4}(b) already reveals that, in this case, the force profile depends on $Pe_m$ in a nonmonotonic fashion (compare the data for $Pe_m=20$ with  the two other data sets).  This behavior is more thoroughly shown for the contact force as a function of $Pe_m$ at different fixed values of $Pe_c$ in Fig. \ref{fig:fig5}(b). The force indeed increases up to an intermediate value of $Pe_m^\ast$, where it takes a pronounced global maximum. For $Pe_c=0$, we find  $Pe_m^\ast\simeq35$, a value that becomes significantly smaller, when $Pe_c$ is fixed at larger values, as shown the figure.   

In the cases where $Pe_c$ is kept fixed (Fig. \ref{fig:fig5}(b)), there are several factors that control (e.g., enhance or suppress) the effective force imparted on the inclusions. By increasing $Pe_m$, the ABPs present in the exterior regions more deeply overlap with the enclosing membranes. The internal concentration of ABPs increases and this is the factor that enhances the effective force. On the other hand, by increasing $Pe_m$, the average time of interaction between external ABPs and the enclosing  membranes decreases and also the ringlike structures of ABPs outside and inside the inclusions are weakened, as  the exterior ABPs become more strongly motile and make the interior ABPs more strongly motile due to steric repulsion between ABPs. These latter factors are the ones that suppress the effective force produced on the inclusions. The exact contribution due to each of these factors cannot be determined as they are inter-related and they are not expected to contribute linearly to the effective force. At fixed $Pe_c=0$, an initial increase in $Pe_m$ gives a dominant increase in the overlapping of external ABPs with the enclosing membranes and their internal concentration, and thus an increase in the effective force, but these factor are eventually overtaken by the factors that create effective force suppression. At fixed $Pe_c=50$, by increasing $Pe_m$ in the range $[0,50]$, where the internal concentration is diluted (see Fig. \ref{fig:fig3}), there appears to be a competition between the increase in the  overlapping of external ABPs with the enclosing membranes and the decrease in the average time of the interaction between external ABPs and the enclosing membranes. As $Pe_m$ is increased, the former factor becomes dominant and, as a result, the effective force increases but, on further increase of $Pe_m$, the latter factor dominates and, hence, the effective force decreases. In the range $Pe_m>50$, the internal ABP concentration increases with $Pe_m$; this, being the dominant factor in this regime of parameters, causes an increase in  both internal and external forces and, as a result, an increase in the effective force. The increase in the effective force continues until a secondary hump appears, as seen in Fig. \ref{fig:fig5}(b) (blue curve). The said increase stops after the internal concentration is saturated, in which case, the decrease in the interaction time becomes dominant again and, thus, the effective force decreases. At $Pe_c=100$, the same mechanisms hold except that the second hump does not appear in the range $Pe_m<200$, because the internal concentration only weakly increases with $Pe_m$.

\subsection{Regimes of attraction and repulsion}
\label{subsec:phase_diag}

Even though our focus in the preceding sections has mainly been on the regime of parameters, where the forces acting on the inclusions are found to be repulsive (positive), there are other regimes, where the forces turn out to be attractive (negative) and would thus tend to bring the inclusions closer together. We run extensive simulations across a wide range of values over the $(Pe_m, Pe_c)$ plane to evaluate the magnitude and the sign of the net effective force acting on the inclusions and their internal and external components. The results are shown for the net  force $F_0$ acting on the inclusions at contact, its internal component  $F_0^{in}$ and its external component $F_0^{ext}$, fulfilling the relation  $F_0=F_0^{in}+F_0^{ext}$, as color-coded density maps in Fig. \ref{fig:fig6}(a), (b) and (c), respectively. As seen in panel a, the net force is repulsive for sufficiently small $Pe_m$ and/or $Pe_c$, with a maximal value (red spot) obtained for $Pe_c$, i.e., when the interior P\'eclet number is zero and $Pe_m^\ast\simeq35$. The net force becomes attractive in the central (triangular) region of the parameter space enclosed by the dashed curve. The figure indicates a minimum value of $Pe_c\simeq50$ and $Pe_m\simeq40$ below which attraction is not possible.  

Fig. \ref{fig:fig6}(b) and (c) show that a net attraction originates in the internal component, $F_0^{in}$, of the force acting on the inclusions, as $F_0^{ext}$ never becomes attractive. Panel b also indicates that $F_0^{in}$ is attractive typically when $Pe_c\gtrsim 60$, unless $Pe_m$ goes to zero, in which case the threshold $Pe_c$ (above which  $F_0^{in}$ becomes attractive) also tends to zero. 

Further insight can be obtained by comparing the regions of attraction and repulsion in Figs. \ref{fig:fig6}(a)-(c) with the regions of the parameter space, where ABPs are found to be depleted from or saturated within the inclusions; see Fig. \ref{fig:fig3}. Hence, it follows immediately that a net repulsion occurs when the ABPs are predominantly depleted from or saturated within the inclusions. 

The signs of the internal and external forces depend on the distribution of ABPs interacting with the enclosing membranes of the inclusions in their interior and exterior regions. Based on our simulation results for the ABP distributions and the resulting force components, we can divide the $(Pe_m, Pe_c)$ plane to three parametric regions I-III, as shown in Fig.  \ref{fig:fig6}(d). The boundary lines here are determined by numerical interpolation of data obtained by fixing $Pe_c$ ($Pe_m$) and scanning the  $Pe_m$ ($Pe_c$) axis at the resolution of $\Delta P_m=5$ ($\Delta P_c=5$). In all of these three parametric regions, ABPs are more strongly concentrated at the narrow and wedge-shaped exterior region, intervening the two inclusions, where ABP trapping effects are predominant. This is the primary source of the all-repulsive external component of the force $F_0^{ext}$. However, as schematically shown in Fig. \ref{fig:fig7},  in region I, the ABPs that are found within the inclusions show stronger steric overlaps with the enclosing membranes at the distal parts of the interior region (the ABPs found at the proximity of the intervening wedge-shaped gap between the inclusions show weaker steric overlaps with the enclosing membranes mainly because the inside ABPs also experience stronger inward-pointing repulsions from the outside ABPs that are accumulated in the outside inter-inclusion gap). As such, the inside ABPs tend to push the inclusions apart, making the internal force, $F_0^{in}$,  repulsive as well. Region I thus corresponds to the situation, where $F_0^{in}, F_0^{ext}$ and $F_0>0$. In regions II and III, the ABPs that are found within the inclusions are more strongly concentrated at the proximity of the inter-inclusion gap and induce an attractive internal force. In region II, because of low concentration or low motility of ABPs inside the inclusions, such an attractive internal force cannot dominate the repulsive force due to external ABPs and, as a result, the total effective force turns out to be repulsive. This region corresponds to the situation, where $F_0^{in}<0$, $F_0^{ext}>0$ (that is, $|F_0^{in}|<|F_0^{ext}|$) and $F_0>0$. In region III, the inside ABPs acquire sufficient concentration and motility to induce attractive internal force to dominate the repulsive external force. This region corresponds to the situation, where the internal area fraction takes intermediate values, approximately in the range of $0.1-0.5$ (see Fig. \ref{fig:fig3}). In region III, $F_0^{in}<0$, $F_0^{ext}>0$ (with $|F_0^{in}|>|F_0^{ext}|$) and $F_0<0$.

\begin{figure}[t!]
\begin{center}
\includegraphics[width=0.75\textwidth]{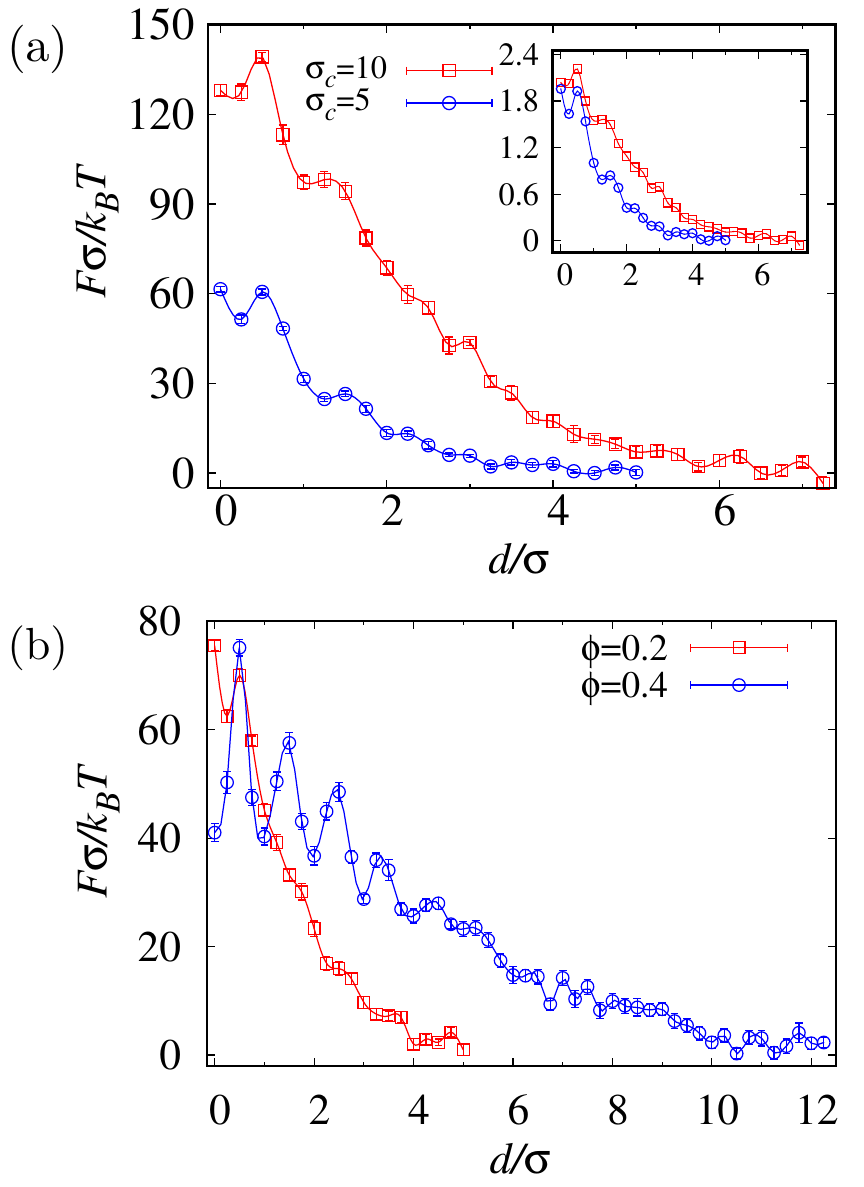}
\caption{(a) Effective force acting on the inclusions as a function of their rescaled surface-to-surface distance, $d/\sigma$, for two different sizes of the inclusions $\sigma_c=5$ and $10$ at fixed overall area fraction of ABPs in the system, $\phi=0.2$, and  area fraction of inclusions, $\phi_c=0.05$. The inset shows the force divided by the perimeter of the inclusions as a function of $d/\sigma$. (b) The same quantity plotted for two different overall area fractions of ABPs in the system $\phi=0.2$ and $0.4$  at fixed area fraction of inclusions, $\phi_c=0.025$. In both panels, we have fixed $Pe_c=0$ and $Pe_m=20$.}
\label{fig:fig8}
\end{center}
\end{figure}

\subsection{Role of inclusion size and ABP area fraction}

The general behavior of the force profiles with the surface-to-surface distance between the inclusions remains the same as the inclusion size is increased; see Fig. \ref{fig:fig8}(a). The larger the inclusion size the larger will be the magnitude and the range of the force experienced from the ABPs. This is expected because the increased perimeter of the enclosing membrane of the inclusions and also its reduced curvature lead to larger residence times for the ABPs near the inclusions and, hence, larger proximal ABP concentrations \cite{Mallory2014}; this is further supported by the effective force divided by the perimeter of inclusions in the inset of Fig. \ref{fig:fig8}(a). 

In systems with larger area fractions, a relatively larger fraction of ABPs accumulate inside and around the inclusions, which in effect leads to a larger and longer-ranged force profile as shown Fig. \ref{fig:fig8}(b). In this case, a larger number of ABP rings form around the inclusions as reflected by the larger number of peaks  in the force profiles. 

\section{Concluding remarks}
\label{sec:conclusion}
We investigate  spatial distribution of ABPs and effective interactions mediated by them between permeable (hollow) dislike inclusions with  mismatching ABP motility (self-propulsion) strengths in their interior/exterior regions. The inclusions are enclosed in permeable membranes modeled using soft, repulsive interfacial potentials.  The ABPs are shown to distribute in ringlike high-density layers inside and outside the inclusions, which  strongly influence the effective interactions mediated between the inclusions through the active bath.  We determine the regimes of effective repulsive/attractive and elucidate their underlying mechanisms  by decomposing the net force between the inclusions to external and internal force components; while the external force turns out to be repulsive, the internal force is found to be  repulsive (attractive) below (above) a certain  threshold interior P\'eclet number,  indicating that a net attraction stems from a dominant internal force. 

Our study builds on a basic model of ABPs interacting only through steric repulsions and demonstrates how a  relatively complex nonequilibrium phenomenology emerges from a minimal description of inclusion permeability and spatial inhomogeneity in ABP motility. Our model can be generalized to include other types of spatial inhomogeneity in the system such as mismatching interior/exterior values for particle diffusivity (temperature), fluid viscosity (especially when hydrodynamic stresses are relevant), and stimulus/field  intensities.  The latter can be realized by subjecting the interior/exterior regions to external stimuli (e.g., activating light \cite{Lozano2016,GomezSolano2017}) or fields (e.g., magnetic/gravitational fields) that impart forces and/or torques of differing magnitudes on the ABPs inside and outside the inclusions. In all such cases, however, particle activity is expected to emerge as a crucial factor. Indeed,  switching off the self-propulsion mechanism  in the present model also leads to complete elimination of the phenomenology reported.

Our model assumes that the inclusions are placed at fixed positions within the active bath, an assumption that has frequently been used in the literature \cite{Harder2014,Naji2017,Lowen2015,Cacciuto2014,Bolhuis2015,Reichhardt2014,Ferreira2016,Zhang2018,Zhang2020, Pagonabarraga2019,Naji2019,Selinger2018,Yang2020,Marconi2018}, as it facilitates  direct analogies with textbook examples of equilibrium depletion forces between colloidal inclusions fixed within a bath of nonactive depletants \cite{Lekkerkerker2011,Likos2001}, where  the bath-mediated interactions  represent  the corresponding potentials of mean force between the inclusions. Several studies involving  mobile (hard) inclusions in an active bath have also been published \cite{Yang2020,Mishra2018, Garcia2015,Harder2014,Leonardo2011,Narayanan2018}. Effective interactions obtained in the case of fixed inclusions are  expected to be different  from those obtained in the case of mobile inclusions in an active  bath in a similar setting  \cite{Yang2020}, and possible relations between the results in the two cases remain to be understood. It is thus interesting to study the role of inclusion mobility in the context of permeable inclusions as well, where many-body effects  \cite{Naji2017} can also be examined by assuming a suspension of many such inclusions within the active bath.

Possible real-world  examples of permeable inclusions may be furnished by fluid enclosures such as lipid and polymer vesicles  \cite{Larsson2006,Kamat2011,Rideau2018}, cells, soft tissues as well as immiscible/stabilized emulsion drops  and active droplets \cite{Cates2017,Tjhung2012,Hyman2014,Julicher2017,Vladescu2014,Naji2018}.  Polymersomes (polymer vesicles) fabricated with controlled permeability \cite{Joseph2017} and polymer stomatocytes derived from controlled deformation of polymersomes \cite{Wilson2012} have recently been used to selectively  entrap catalytically active platinum nanoparticles. In the biological context, cells can ingest and internalize foreign particles through phagocytosis; a process recently utilized in controlled internalization and cell-assisted assembly of nonactive polystyrene microparticles and crystallites within  fibroblast cells \cite{Kodali2007}. Our study should motivate further studies along these lines using active particles as phagocytosed components.  Active particles have also emerged as efficient agents for targeted drug and cargo delivery \cite{Peng2018,Joseph2017,Schmidt2018,Sitti2019,Magdanz2019,Ghosh2020}. Soft biological tissues such as tumors can exhibit enhanced permeation to suitably designed active particles such as platinum-sputtered polymersome nanomotors \cite{Peng2018}, and  biological microswimmers such as magneto-aerotactic bacteria \cite{Felfoul2016,Ghosh2020}.

Recent studies of semipermeable vesicles filled with active particles \cite{Tian2017, Chen2017, Shan2019, Wang2019, Paoluzzi2016, Spellings2015,Naji2018} can also be generalized straightforwardly to incorporate permeation of active particles through the enclosing vesicle membrane  \cite{Daddi2019a,Daddi2019b}. 
Such extensions can provide  insight into the interplay between permeability and shape fluctuations/deformability, brining the model vesicles closer to realistic examples of soft permeable inclusions. 

\section{Conflicts of interest}

There are no conflicts of interest to declare. 

\section{Acknowledgements}
A.N. acknowledges support from the Associateship Scheme of The Abdus Salam International Centre for Theoretical Physics (Italy). We thank M. Zarif and S. Abbasi for useful comments, and acknowledge support from the High Performance Computing Center, IPM.

\section{Author contributions}
M.S. developed the numerical simulations, analyzed the output data and generated the figures. Both authors contributed to the discussions and development of the concepts presented herein, and wrote the manuscript. A.N. conceived the study and supervised the research.

\end{document}